\documentclass[twocolumn,prb,superscriptaddress,showpacs,footinbib]{revtex4}

\usepackage[utf8]{inputenc}
\usepackage[T1]{fontenc}
\usepackage{times}
\usepackage[american]{babel}
\usepackage{graphicx}
\usepackage{xcolor}
\usepackage{booktabs}
\usepackage{amsmath, amssymb}
\usepackage{braket}
\usepackage{xspace}
\usepackage[unicode]{hyperref}

\renewcommand{\vec}{\boldsymbol}

\newcommand{\avr}[1]{\braket{#1}}			
\newcommand{\abs}[1]{\left|#1\right|}			

\newcommand{\imag}{\mathrm{i}}				

\newcommand{\td}[1][]{\mathrm{d}^{#1}}			
\newcommand{\pd}[1][]{\partial^{#1}}			
\newcommand{\tdfrac}[3][]{\frac{\td[#1]#2}{\td{#3}^{#1}}}
\newcommand{\pdfrac}[3][]{\frac{\pd[#1]#2}{\pd{#3}^{#1}}}

\makeatletter
\newcommand{\db}[1][]{\td[#1]\protect\@latex@warning{deprecated command -- use "\td" instead}}
\newcommand{\pdb}[1][]{\pd[#1]\protect\@latex@warning{deprecated command -- use "\pd" instead}}
\newcommand{\ddb}[3][]{\tdfrac[#1]{#2}{#3}\protect\@latex@warning{deprecated command -- use "\ddfrac" instead}}
\newcommand{\pddb}[3][]{\pdfrac[#1]{#2}{#3}\protect\@latex@warning{deprecated command -- use "\pdfrac" instead}}
\makeatother

\newcommand{\elon}{\text{e}}		
\newcommand{\hole}{\text{h}}		
\newcommand{\exc}{\text{X}}		

\newcommand{\Coul}{\text{C}}		
\newcommand{\Hart}{\text{Ha}}		

\newcommand{\inplane}{\shortparallel}	

\newcommand{\quantity}[1]{\ensuremath{#1}\xspace}

\newcommand{\kB}{\quantity{k_{\text{B}}}}		

\newcommand{\aB}[1][]{\quantity{a_{\text{B}#1}}}	
\newcommand{\rs}[1][]{\quantity{r_{\text{s}#1}}}	
\newcommand{\gs}[1][]{\quantity{\gamma_{\text{s}#1}}}	
\newcommand{\qdeg}{\quantity{\chi}}			

\newcommand{\ie}{{i.e.}\xspace}
\newcommand{\eg}{{e.g.}\xspace}

\begin{document}
\title{Crystallization of an exciton superfluid}
\date{\today}

\author{J.~Böning, A.~Filinov and M.~Bonitz}
\email{bonitz@physik.uni-kiel.de, filinov@theo-physik.uni-kiel.de}
\affiliation{Institut für Theoretische Physik und Astrophysik,
Christian-Albrechts-Universität, Leibnizstr. 15, D-24098 Kiel, Germany}

\begin{abstract}
Indirect excitons -- pairs of electrons and holes spatially separated in semiconductor bilayers or quantum wells -- are known to undergo Bose-Einstein condensation and to form a quantum fluid. Here we show that this superfluid may crystallize upon compression. However, further compression results in quantum melting back to a superfluid. This unusual behavior is explained by the effective interaction potential between indirect excitons which strongly deviates from a dipole potential at small distances due to many-particle and quantum effects. Based on first principle path integral Monte Carlo simulations, we compute the complete phase diagram of this system and predict the relevant parameters necessary to experimentally observe exciton crystallization in semiconductor quantum wells.
\end{abstract}

\pacs{71.35.Lk, 03.75.Hh, 05.30.Jp}

\maketitle

\section{Introduction}

Quantum coherence of bosonic particles is one of the most striking macroscopic manifestations of the laws of quantum mechanics governing the microworld. The discovery of Bose-Einstein condensation in atomic vapors~\cite{anderson_observation_1995} was followed by the observation of condensation of bosonic quasiparticles in condensed matter -- excitons. Here we mention early claims (though highly controversial) for three-dimensional (3D) semiconductors,~\cite{butov_towards_2002} electron bilayers in a quantizing magnetic field,~\cite{macdonald_fractional_1990, tiemann_critical_2008} exciton-polaritons in microcavities~\cite{kasprzak_bose-einstein_2006, amo_superfluidity_2009} and so-called indirect excitons formed from spatially separated electrons and holes.~\cite{lozovik97,filinov_excitonic_2003, filinov_path_2006, timofeev_collective_2007, ludwig_quantum_2006}
Not only the bosonic gas phase was observed but also the formation of a quantum Bose liquid -- an exciton superfluid with its peculiar loss of friction -- could recently be verified.~\cite{tiemann_critical_2008, amo_superfluidity_2009} Thus it is tempting to ask whether there exists also a solid phase of bosons.

The key properties of a crystal are particle localization and long-range spatial ordering.
To achieve spontaneous crystallization requires to find a Bose system with sufficiently strong and long range pair interaction (here we do not consider particle localization induced by an external field in an optical lattice or cavity~\cite{domokos_collective_2002, gopalakrishnan_emergent_2009}). However, the vast majority of previous experimental investigations
have been performed in the regime of weak nonideality,  where the interaction energy is small compared to the quantum kinetic energy.
Therefore, promising candidates for a bosonic solid are atoms or molecules with dipole interaction~\cite{griesmaier_bose-einstein_2005} or excitons.
Here, indirect excitons offer a number of attractive features: a strong dipole-type interaction, the suppression of biexciton or trion formation, the comparatively long radiative life time (on the order of microseconds) and the external controllability of the density and dipole moment via an electric field perpendicular to the quantum well plane.~\cite{ludwig_quantum_2006, timofeev_collective_2007, sperlich_electric_2009}

In this paper we present clear evidence for the existence of a crystal of indirect excitons in semiconductor quantum wells. We compute its full phase diagram and reveal the parameters for its experimental verification. Our predictions are based on first principle path integral Monte Carlo (PIMC) simulations. But in contrast to previous quantum Monte Carlo studies which predicted crystallization in model systems such as electron-hole bilayers, ~\cite{de_palo_excitonic_2002, filinov_excitonic_2003} or two-dimensional dipole systems,~\cite{astrakharchik_quantum_2007, filinov_berezinskii-kosterlitz-thouless_2010} here we use realistic parameters typical for indirect excitons. In particular, we fully take into account the finite quantum well width, the composite character of the excitons and the different masses of electrons and holes. This turns out to be of crucial importance for the exciton-exciton interaction which strongly departs from a dipole potential at small distances. As a direct consequence we observe that the exciton crystal exists only in a finite density interval and undergoes quantum melting both at high and low density. Furthermore -- when the exciton superfluid crystallizes to form a solid, quantum coherence is lost abruptly, \ie there is no supersolid exciton phase.

This paper is organized as follows. In Sec.~\ref{sec_model} we introduce the system of indirect excitons and
present its reduced quasi-2D description. In Sec.~\ref{ex_pot} the effective exciton-exciton interaction potential is derived and its accuracy is verified. In Sec.~\ref{results} we present our simulation results and the phase diagram of indirect excitons. Finally, we draw our conclusions in Sec.~\ref{final}.

\section{Model}\label{sec_model}

We consider a semiconductor quantum well (QW) of width $L$ containing $N_\elon=N_\hole$ electrons and holes in the conduction and valence band, respectively, which are created by an optical pulse.~\cite{note1} Application of an electrostatic field of strength $E$ perpendicular to the QW plane created \eg by a tip electrode allows to spatially separate electrons and holes to different edges of the QW. By varying $E$ this separation can be changed between $0$ and $L$ giving rise to a variable dipole moment $d$. At the same time, the field also provides lateral confinement and a variable particle density, via the quantum confined Stark effect, for details of the setup, see K.~Sperlich \emph{et al}.~\cite{sperlich_electric_2009} Finally, the system is kept in thermal equilibrium at a finite temperature $T$ which does not exceed a few percent of the binding energy of an electron-hole pair, thus all electrons and holes will be bound in $N=N_\elon$ indirect excitons.~\cite{note2}

The thermodynamic properties of this system are fully described by the density operator of $N_\elon$ electrons and $N_\hole$ holes, ${\hat \rho}^A_{N_\elon,N_\hole} = {\cal A}\{e^{-\beta {\hat H}}/Z\}$, where $Z$ is the partition function, $\beta=1/(\kB T)$ and ${\cal A}$ denotes full anti-symmetrization among all electronic and hole variables. The full Hamiltonian ${\hat H}$ contains kinetic energy, the interaction with the external electric field and all Coulomb pair interactions between the $2N_\elon$ charged particles
\begin{align}
\hat H &= \hat H_{\parallel}+\hat H_z+W\;,
\label{H1}
\end{align}
with the single particle contributions
\begin{align}
\hat H_{\parallel} &= \sum_{i=1}^{N} \left[ -\frac{\hbar^2 \nabla_{\vec{r}_i}^2}{2 m^{\parallel}_\text{e(h)}} \right]\;, \nonumber\\
\hat H_z &= \sum_{i=1}^{N} \left[ -\frac{\hbar^2 \nabla_{z_i}^2}{2 m^{\perp}_\text{e(h)}}  + V^\text{QW}_\text{e(h)}(z_i)+U_\text{e(h)}\{E_z\}\right]\;,
\end{align}
and the interaction part
\begin{align}
W &= \sum_{i<j}^{N} V^\text{Coul}_{ij}\;, &
V^\text{Coul}_{ij} &= \frac{e_i e_j}{\epsilon\sqrt{r_{ij}^2+z_{ij}^2}}\;.
\end{align}
Here $r_{ij}$ denotes inter-particle distances in the QW plane, $V^{QW}$ is the QW confinement,  $U$ is the electrostatic potential energy due to the electric field and $\epsilon$ is the background dielectric constant, $m^{\parallel}$ and $m^{\perp}$ are the the effective electron (hole) masses which
take into account the anisotropy of the in-plane (out of-plane) parabolic dispersions in the QW.

Under the present conditions of strongly bound indirect excitons with parallel dipole moments resulting in a strong exciton-exciton repulsion the very complicated evaluation of the density operator $\hat \rho^\text{A}_{N_\elon,N_\hole}$ can be substantially simplified. As was shown in Ref.~\cite{filinov_effective_2009} the system can be mapped onto $N$ excitons which can be treated as composite spin polarized bosons~\cite{note3} where deviations from the Bose statistics (arising from the original Fermi statistics of electrons and holes) have been found negligible.~\cite{filinov_path_2006} Thus, the density operator is reduced to a fully symmetric one of $N$ excitons, $\hat\rho^\text{S}_{N}$. Furthermore, all pair interactions can be properly averaged along the QW width giving rise to an effective ($d$-dependent) exciton-exciton interaction $V_{\exc\exc}$. As a result the system 2D Hamiltonian entering $\hat\rho^\text{S}_{N}$ becomes
\begin{align}
  \hat H^\text{eff} = \sum_{i=1}^{N} \left[ -\frac{\hbar^2 \nabla_{\vec r_i}^2}{2m_\exc} \right]
    + \sum_{i<j} V_{\exc\exc}(r_{ij};d)\;,
  \label{eq:Heffexc}
\end{align}
where $m_\exc=m_\elon^\inplane+m_\hole^\inplane$ is the in-plane effective mass, $\vec r_i$ the in-plane center of mass (com) coordinate of the $i$\textsuperscript{th} exciton and $r_{ij}=\abs{\vec r_i - \vec r_j}$ denotes the com distance between two excitons.

\section{Effective inter-exciton interaction}\label{ex_pot}


To verify the approximation (\ref{eq:Heffexc}) and the validity of the potential $V_\text{xx}$ we consider the two-exciton (biexciton) problem.
We define the exciton interaction energy as the energy difference of a biexciton and two single excitons, $E_\text{XX}(r_\text{hh})=E_\text{2X}(r_\text{hh})-2E_\text{X}$, which depends parametrically on the distance between the holes in a biexciton problem, $r_\text{hh}=\abs{\vec{R}_1-\vec{R}_2}$. The distance $r_\text{hh}$ remains a well defined quantity also at small inter-exciton separations, when a strong overlap of the exciton wavefunctions and particle exchange takes place. In this case the com distance is not physical. The substitution of $r_{ij}$ in Eq.~(\ref{eq:Heffexc}) by  $r_\text{hh}$ can be justified as follows.


Similar to the hydrogen problem, the single exciton wave function can be factorized into the com and the relative part
\begin{align}
 \Psi(\vec{r},\vec{R})&=\Psi_\text{C}(\vec{R}^0)\,\Psi_\text{r}(|\vec{r}-\vec{R}|)\;,
\end{align}
with
\begin{align}
 \vec{R}^0 &= \frac{m_\text{e}}{m_\text{X}}\vec{r}+\frac{m_\text{h}}{m_\text{X}}\vec{R}\;, &
 m_\text{X} &= m_\text{e}+m_\text{h}\;,
\label{eq1}
\end{align}
where the vectors $\vec{r}$, $\vec{R}$ and $\vec{R}^0$ denote the electron, hole and com coordinates, respectively.

The relative part $\Psi_\text{r}$ can be found by solving a single particle problem with the reduced mass $\mu= m_\text{e} m_\text{h}/(m_\text{e}+m_\text{h})$ in the potential, $V_d=-e^2/\sqrt{|\vec{r}-\vec{R}|+d^2}$, where the $z$-direction is taken into account explicitly by the exciton dipole moment $d$. For the spatially indirect exciton we approximate
\begin{align}
 V_d|_{r<d} &= -\frac{e^2}{\sqrt{r^2+d^2}}\eqsim -\frac{e^2}{d}(1-\frac{r^2}{2 d^2} + \ldots)\;,
\end{align}
\begin{align}
 \Psi_\text{r}^H(r)|_{r<d} &\propto e^{-r^2/2 l^2}\;,&
 l^2 &= \frac{\hbar}{\mu \omega}\;, &
 \omega^2 &=\frac{e^2}{d^3}\;,
\label{eq2}
\end{align}
i.e. the leading term of the expansion describes a harmonic oscillator and the relative part near the exciton origin decays as a Gaussian.
Now, using the definition of $\vec{R}_0$ and the substitution, $(\vec{r}-\vec{R})=\gamma_m(\vec{R}^0-\vec{R})$ with $\gamma_m=m_\text{X}/m_\text{e}$, the relative part can be expressed solely in terms of the hole coordinate (keeping the com coordinate $\vec{R}^0$ as a fixed parameter)
\begin{align}
 \Psi(\vec{r},\vec{R}) &= \Psi_\text{C}(\vec{r},\vec{R})\,\Psi_\text{r}(\vec{R},\vec{R}^0)\;,
\label{1ex}
\end{align}
where the relative part~(\ref{eq2}) contains a factor $\gamma_m^2$ in the exponent, $\Psi_\text{r}^H(r)|_{r=|\vec{R}_0-\vec{R}|}\propto  e^{-\gamma_m^2 r^2/l^2}$. For a typical electron-hole mass ratio in semiconductors, $\gamma_m\sim 2\ldots 4$, we conclude, that the hole is well localized around the com. This allows to make a second step.

We treat the excitons in the Born-Oppenheimer (BO) approximation and apply the adiabatic transformation for the spatial part of the full wavefunction~\{Spin degree of freedom are omitted in the present analysis, as this requires a significantly more elaborated simulations.
The model used for the exciton interaction potential, is assumed to have a significantly larger effect on the results, when the spin fluctuations in the  ferromagnetic phase~\cite{note4}
\begin{multline}
 \Psi_{\text{XX}}=\frac{1}{(2!)^2} \sum_{P_\text{e},P_\text{h}} (\pm 1)^{\delta P_\text{e} +\delta P_\text{h}} \Psi_\text{e}(\hat{P}_\text{e}\vec{r}_1,\hat{P}_\text{e}\vec{r}_2,\vec{R}_1,\vec{R}_2)\\
 \times \Psi_\text{h}(\hat{P}_\text{h}\vec{R}_1,\hat{P}_\text{h}\vec{R}_2)\;,
\label{total}
\end{multline}
which can be symmetric or antisymmetric depending on the symmetry of the spin part.
The action of the electron and hole permutation operators, $\hat{P}_\text{e(h)}$, explore all exchange possibilities (excluding the electron-hole exchange). Within this ansatz one can self-consistently solve the Schrödinger equations for electrons
\begin{multline}
 \hat{H}_\text{e} \Psi_\text{e}^{(n)}(\vec{r}_1,\vec{r}_2,\vec{R}_1,\vec{R}_2)\\
 =E_\text{e}^{(n)}(R_{12}) \Psi_\text{e}^{(n)}(\vec{r}_1,\vec{r}_2,\vec{R}_1,\vec{R}_2)\;,
\label{1Sch}
\end{multline}
and holes
\begin{multline}
[\hat{H}_\text{h}+ E_\text{e}^{(n)}(R_{12})]\,\Psi_\text{h}^{(m)}(\vec{R}_1,\vec{R}_2,\vec{R}_1^0,\vec{R}_2^0)\\
 =E_\text{2X}^{(m)} \Psi_\text{h}^{(m)}(\vec{R}_1,\vec{R}_2,\vec{R}_1^0,\vec{R}_2^0)\;,\label{2Sch}
\end{multline}
where
\begin{align}
\hat{H}_\text{e} &= \sum_{i=1,2}
  \left[\hat{T}_\text{e}^i+ V_{\text{eh}}(\vec{r}_i-\vec{R}_1)
    +V_{\text{eh}}(\vec{r}_i-\vec{R}_2)\right]\nonumber\\
  &+V_{\text{ee}}(\vec{r}_1-\vec{r}_2)\;,\\
 \hat{H}_\text{h} &= \sum_{j=1,2} \hat{T}_\text{h}^j+V_{\text{hh}}(R_{12})\;,\\ \hat{T}_\text{e(h)}
  &= -\frac{\hbar^2 \nabla^2}{2 m_\text{e(h)}}\;,
\end{align}
with $n,m\in\set{A,S}$ being defined by the symmetry of the electron (hole) wavefunction, and $E_\text{e}^{(n)}$ being an additional mean-field electron potential influenced by the holes in the biexciton.

If the holes are treated as infinitely heavy,~\cite{Zim} the numerical solution of Eq.~(\ref{2Sch}) is not necessary and the biexciton energy can be decomposed, $E_\text{2X}=E^{(n)}_{\text{e}}(r_{\text{hh}})+V_{\text{hh}}$, with $V_{\text{hh}}=e^2/\abs{\vec{r}_{\text{hh}}}$. The electron contribution $E^{(n)}_{\text{e}}$ is the solution for a singlet (triplet) state
\begin{multline}
\left[\sum_{i=1,2} \left(-\frac{\hbar^2\nabla^2_{r_i}}{2 m_\text{e}}
  + V_\text{eh}(\vec{r}_i)\right)+ \frac{e^2}{|\vec{r}_1-\vec{r}_2|}\right]\,
  \Psi_\text{e}^{S/A}\\
= \left[ E_\text{e}^{S/A}+2E(\text{X}) \right]\,\Psi_\text{e}^{S/A}\;,
\end{multline}
where
\begin{align}
V_{\text{eh}}(\vec{r}_i) &= \sum_{j=1}^2 -\frac{e^2}{\sqrt{(\vec{r}_i+ \vec{R}_j)}+d^2}\;,
\label{Scheq}
\end{align}
with the holes located at $\vec{R}_{1,2}=\pm \frac{1}{2}\vec{r}_{{\text{hh}}}$.
%
\begin{figure}
\includegraphics{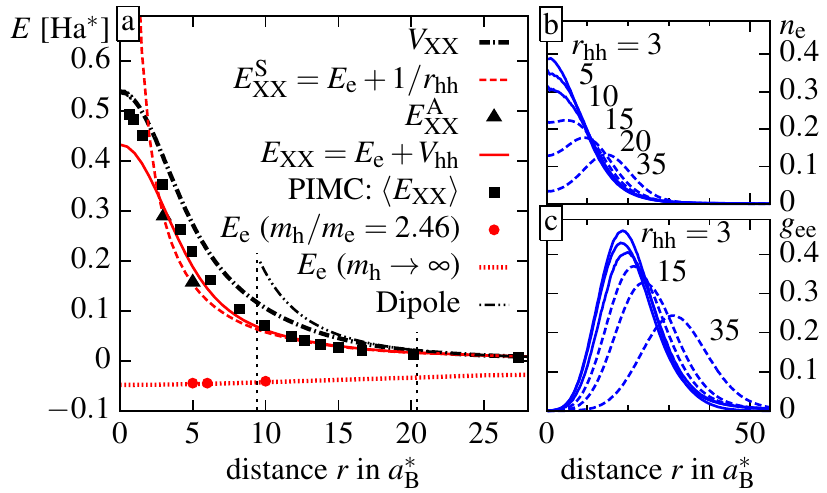}
\caption{(Color online) Exciton interaction potential $V_{\text{XX}}$ for a dipole moment $d=13.3 \aB^*$.
(a) The interaction potential $V_{\text{XX}}$ (point-dashed line) is compared to the exciton interaction energy $E_{\text{XX}}$ in several approximations: average interaction of two excitons $\langle E_{\text{XX}} \rangle$ evaluated by PIMC simulations using the Hamiltonian, Eq.~(\ref{eq:Heffexc}); the BO-model with the infinite hole mass, symmetric ($E_{\text{XX}}^S)$ and antisymmetric ($E_{\text{XX}}^A$) electronic states; the improved BO-model, $E_{\text{XX}}=E_{\text{e}}+V_{\text{hh}}$, with a realistic mass ratio $m_\text{h}/m_\text{e}=2.46$ (ZnSe-based QW). Also shown are the electronic contribution $E_{\text{e}}$  and the dipole potential $d^2/r^3$. Two vertical lines indicate the boundaries of the exciton crystal.
(b) Radial electron density $n_{\text{e}}(r)$ for several hole separations $r_{\text{hh}}$, relative to the mid-point of two holes located at $\vec{R}_{1,2}=\pm \frac{1}{2}  \vec{r}_{\text{hh}}$. (c)  Electron pair distribution function
$g_{\text{ee}}(r_{\text{ee}})$ for the values $r_{\text{hh}}$ in (b).}
\label{fig:effpot}
\end{figure}
This equation has been solved numerically for an experimentally feasible \elon-\hole\ separation $d=13.3 \aB^*$.~\cite{note5}
A first observation is that, the energy $E_\text{e}(m_\text{h}\to\infty)$ is not sensitive to $r_\text{hh}$,
once $r_{{\text{hh}}} \lesssim d$, see Fig.~\ref{fig:effpot}a. This is understood from the behavior of the electron density (see Fig.~\ref{fig:effpot}b): in all cases the electron cloud extends well beyond $r_\text{hh}$ which is a result of the shallow interaction potential, $V_{eh}(r)$, of an electron with the two holes for $r_{\text{hh}} < d$, and the strong e-e repulsion that keeps the electrons at an average distance $\tilde r \sim 20 \aB^*$ apart, practically independent on the hole-hole separation.
This behavior is evident from the pair distribution function $g(r_{\text{ee}})$, see Fig.~\ref{fig:effpot}c. Consequently, for a large exciton dipole moment, we observe no noticeable difference in the energy of the symmetric and antisymmetric states, merging into a single curve $E_\text{e}(m_{\text{h}}\rightarrow \infty)$, see  Fig.1a.
With these results we can now analyze $E_{\text{XX}}(r_\text{hh})$, cf. red dashed line in Fig.~\ref{fig:effpot}a. At large distances, $r \gtrsim d$, $E_{\text{XX}}$ practically coincides with the classical dipole potential,~$V_\text{D}=d^2/r^3$, so we expect the system to behave like 2D polarized dipoles, at low densities. At smaller distances, $r<d$, however, $E_{\text{XX}}$ essentially follows a Coulomb potential which arises mainly from the hole-hole repulsion. Finally, for $r \ll d$, the interaction energy shows an unphysical Coulomb singularity originating from the assumption of an infinite hole mass.
In real systems, $E_{\text{XX}}$ is expected to be softer, approaching a finite value at zero distance, due to quantum diffraction and exchange effect, similar to behavior of the Kelbg potential in 3D electron-ion plasmas.~\cite{kelbg,pre0_filinov,pre_filinov} Therefore, we proceed with the generalization of the model for a finite hole mass.

In the situation with a large dipole moment, as considered in Fig.~\ref{fig:effpot}a, the interaction energy is positive at all distances and, hence, no bound states (biexcitons) are formed.  This originates from the positive eigenvalues of the Schrödinger equation for the holes~(\ref{2Sch}). Therefore, evaluation of the interaction energy should not be limited only to the ground state solution of Eq.~(\ref{2Sch}), but should include contribution of all states, including the continuum.~\cite{note6} This can be done directly via the two-particle partition function $Z_2$,
\begin{multline}
Z_2(\beta, r_\text{hh})=\int \td \vec{R}_1  \td \vec{R}_2 \, \rho(\vec{R}_1,\vec{R}_2;\vec{R}_1,\vec{R}_2;\beta)\\
 \times \delta \left(\abs{\vec{R}_1-\vec{R}_2}-r_\text{hh}\right)\;,
\label{z1}
\end{multline}
the density matrix, and the thermodynamic energy estimator
\begin{align}
E(r_\text{hh}) &= -\frac{\partial}{\partial \beta} \ln Z_2(\beta, r_\text{hh})\;.
\label{z2}
\end{align}
Here, $Z_2$ parametrically depends on the distance $r_\text{hh}$ between the particles. Applied to the case of two holes in the biexciton ($E\equiv E_\text{2X}$), the density matrix is the solution of the two-body Bloch equation with the Hamiltonian, $\hat{H}_\text{h}+ E_\text{e}^{(n)}(\abs{\vec{R}_1-\vec{R}_2})$, see Eq.~(\ref{2Sch}), which can be factorized into the com free particle density matrix and the relative part
\begin{align}
 \rho(\vec{R}_1,\vec{R}_2;\vec{R}_1',\vec{R}_2';\beta)&=\rho_F(\vec{R}_c,\vec{R}_c';\beta)\, \rho(\vec{r}_\text{hh},\vec{r}_\text{hh}';\beta)\label{u1}\;,
\end{align}
where
\begin{align}
 \rho(\vec{r}_\text{hh},\vec{r}_\text{hh}';\beta) &\equiv \rho_F(\vec{r}_\text{hh},\vec{r}_\text{hh}';\beta) \, e^{-U^\text{eff}(\vec{r}_\text{hh},\vec{r}_\text{hh}';\beta)}\;.
\label{u2}
\end{align}
Here $U^\text{eff}$ is the effective pair action,~\cite{kelbg,pre_filinov} introduced in a way that at large distances and (or) high temperatures it reduces to $\beta(e^2/\abs{\vec{r}_{\text{hh}}}+E_\text{e}^{(n)}(r_\text{hh}))$. Substituted in Eq.~(\ref{z1})-(\ref{z2}) we obtain
\begin{align}
 E_\text{2X}(r_\text{hh};\beta) &= \kB T+\left(\kB T+\frac{\partial}{\partial \beta} U^\text{eff}(\vec{r}_\text{hh},\vec{r}_\text{hh};\beta)\right),
\end{align}
where the first term accounts for the com kinetic energy (in 2D). For spherically symmetric potentials the effective action and its
temperature derivative can be evaluated with the matrix-squaring technique.~\cite{Storer,Storer2} The resulting interaction energy, $E_{\text{XX}}(r_\text{hh};\beta)=E_\text{2X}(r_\text{hh};\beta)-E_\text{X}(\beta)$, evaluated at the temperature $1/\beta=10^{-3}$Ha$^{*}$ is shown in Fig.~1a by the red solid line. Quantum effects arising from the finite hole mass (e.g. for the ZnSe-based QWs, $m_\text{h}/m_\text{e}\simeq 2.46$) strongly affect the interaction energy $E_\text{XX}$ for $r < 3 \aB^{*}$, which consequently approaches a finite value at zero distance.

For final comparison, we compute the exciton interaction energy by PIMC simulations using the Hamiltonian~(\ref{eq:Heffexc}). We used two bosonic excitons of mass $m_\text{X}$ in periodic boundary conditions. The result, $\langle E_{\text{XX}} \rangle$, as a function of the average inter-exciton distance, $\langle r \rangle=\int \td \vec{r} \, r  g(r)  \cdot (\int \td \vec{r} \,g(r))^{-1}$, evaluated via the exciton pair distribution function $g(r)$, is shown in Fig.~1a by the solid squares. This quantity agrees well with the finite-mass BO solution, $E_{\text{XX}}$, for $r_\text{hh}> 5 \aB^*$, and confirms applicability of both models in the density range where we predict formation of the excitonic crystal. The deviations being noticeable at smaller distances are outside the density range used in the present analysis.


\section{Simulation results}\label{results}

Using PIMC simulations with ${\hat \rho}^s_{N}$ and the Hamiltonian (\ref{eq:Heffexc}) the thermodynamic properties of the $N$ strongly correlated  excitons can be efficiently computed with full account of all interactions, quantum and spin effects, without further approximations.
Below we use atomic units, \ie lengths will be given in units of the electron Bohr radius,
$\aB^* = \hbar^2\epsilon/(e^2m_\elon^\inplane)$, and energies in units of the electron Hartree, $\Hart^*=e^2/(\epsilon\aB^*)$.
Of central importance for the crystallization is the coupling (nonideality) parameter, \ie the ratio of interaction energy to kinetic energy. For a quantum system with Coulomb (dipole) interaction it is given by the Brueckner parameter $\rs$ (the dipole  coupling parameter $D$),
\begin{align}
  \rs&\equiv\frac{a}{\aB^*}\sim n^{-1/2}\;, &
D &\equiv
    \frac{M_\exc}{m_\elon^\inplane}\,\frac{1}{\sqrt\pi\rs}\,\frac{d^2}{\aB^{*2}}\sim n^{1/2}\;,
\end{align}
where $a$ is the mean inter-particle distance and $n$ the exciton density. Note the opposite scaling of $\rs$ and $D$ with
density.

We perform 2D grandcanonical PIMC simulations~\cite{boninsegni_worm_2006} with periodic boundary conditions and extract the results for the canonical ensemble with $N=60\dots 500$ excitons. To map out the phase diagram we scan a broad parameter range spanning three orders of magnitude of density and temperature. We first obtain the phase diagram for a fixed value of the dipole moment, corresponding to $d=13.3\,\aB^*$, and after that analyze in Sec.~\ref{var_d} how the crystal phase boundary changes when $d$ is varied.
\begin{figure}
\includegraphics{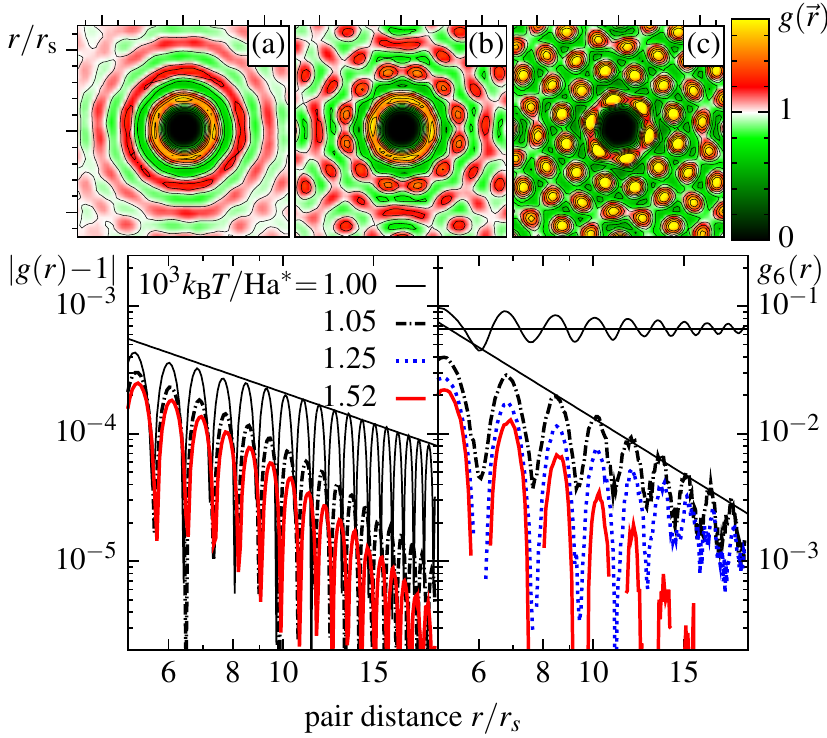}
\caption{(Color online) Constant density freezing. Top row:  2D PDF $g(\vec r_{ij})$ [relative to a fixed particle in the center] for $n\aB^{*2}=0.0035$ and temperatures $\kB T/\Hart^*$ of $1.74\cdot10^{-3}$ (a), $1.38\cdot10^{-3}$ (b) and $1.08\cdot10^{-3}$ (c). Bottom: Radial distribution function $\abs{g(r)-1}$ (left) and bond angular order distribution function $g_6(r)$ (right) at $n\aB^{*2}=0.0022$ for $N=500$.
Lines are guide to the eye to visualize an algebraic decay in this log-log plot.
}
\label{fig:melttemp}
\end{figure}

\subsection{Spatial ordering of excitons}

To detect crystallization we compute the exciton pair distribution function [PDF], $g(\vec{r})$. This function is homogeneous in an ideal gas, whereas in the fluid and crystal phase it exhibits increasing modulations which signal localization and spatial ordering. Typical examples of $g(\vec{r})$ are displayed in the top rows of Figs.~\ref{fig:melttemp} and \ref{fig:meltdens} and show clear evidence of exciton localization. The existence of the translational long range order (LRO) is detected from the asymptotic behavior of the angle-averaged function $g(r)$ for large $r=\abs{\vec r}$. In 2D a possible freezing scenario is given by the Kosterlitz-Thouless-Nelson-Halperin-Young (KTNHY) theory (see the overview~\cite{strandburg_two-dimensional_1988}), predicting an exponential (algebraic) decay of the peak heights of $g(r)$ above (below) the melting temperature. Indeed, our simulations find some support for this scenario, see bottom left part of Fig.~\ref{fig:melttemp}.

\begin{figure}[t]
\begin{center}
\includegraphics[width=0.45\textwidth]{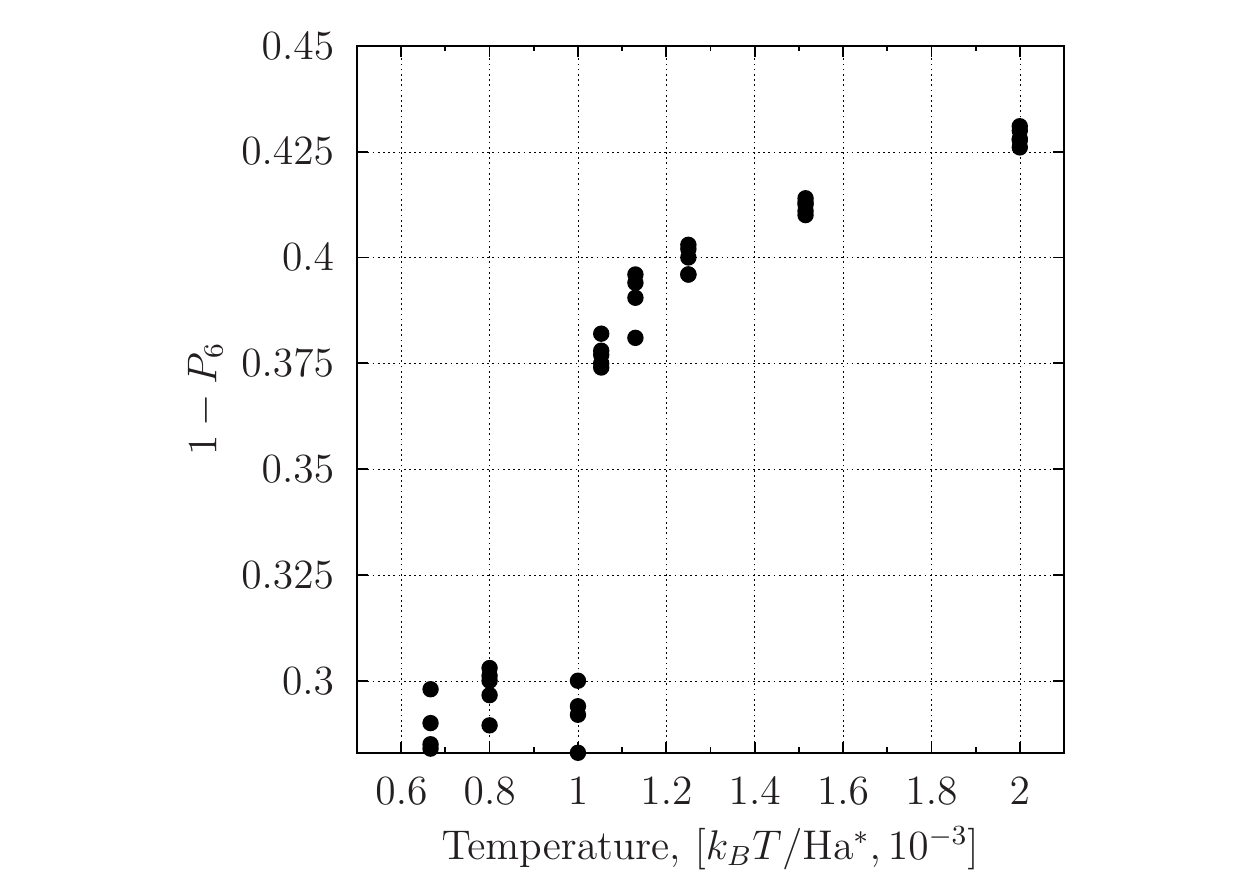}
\end{center}
\vspace{-0.2cm}
\caption{Temperature dependence of the defect fraction at density $n \aB^{*2}=2.2 \cdot 10^{-3}$ and particle numbers $N=501-505$ (vertically aligned dots) characterizing the ordered and disordered phases. The defect fraction  ($1-P_6$) shows a sharp jump at
$T \sim 10^{-3}$Ha$^*$. This is in disagreement with the KTNHY theory, which predicts a continuous unbinding of dislocations in the hexatic phase indicated by a continuous variation of the critical exponent  $\eta_6(T) \leq 1/4$ and the bond angular correlation function $g_6(r)\sim r^{-\eta_6(T)}$. In contrast, we observe an abrupt transition from the LR angular order ($T< T_c$) to a quasi-long-range angular order with $\eta_6(T)\sim 2$ ($T\geq T_c$), see Fig.~\ref{fig:melttemp}.}
\label{figs2}
\end{figure}

The existence of angular hexagonal LRO follows from the asymptotic behavior of the bond angular correlation function, $g_6(r)=\avr{\psi_6^*(r)\psi_6(0)}$, with $\psi_6(r_k)=n_l^{-1}\sum_{l=1}^{n_l}e^{\imag 6\Theta_{kl}}$, where $n_l$ is the number of nearest neighbors of a particle located at $r_k$, and $\Theta_{kl}$ is their angular distance. We observe a change from an exponential asymptotic of $g_6$ to a constant which is the expected behavior for a liquid-solid transition, see bottom right part of Fig.~\ref{fig:melttemp}. There are some indications for the existence of an hexatic phase -- coexistence of angular quasi-LRO (algebraic decay) and missing translational LRO in a narrow temperature interval, see curves for $\kB T=1.05 \cdot 10^{-3}\,\Hart^*$ and $\kB T=1.25\cdot 10^{-3}\,\Hart^*$.

In addition we performed a Voronoi analysis, which provides access to local distortions of the hexagonal symmetry of the lattice.
The average fraction of particles (the probability) with a number of nearest neighbors deviating from $6$ is referred to as the defect fraction, i.e. $(1-P_6)$. The results of Fig.~\ref{figs2} explore the nature of the melting transition at constant density. We observe a sharp increase of the number of defects at the melting point which is in disagreement with the KTNHY scenario. A possible alternative to the KTNHY is a first order solid-liquid phase transition, with an exponential decay of $g_6(r)$. However, the latter was not observed in our simulations, possibly, due to a limited system size ($N \sim 500$). The constructed Voronoi map for different particle configurations, shows the accumulation of the defects at the boundaries between few crystallites. A similar picture, but for a significantly larger classical system ($N\sim 10^{6}$) has been recently reported and the transition was proved to be of the first order.~\cite{Hart} If that system was equilibrated sufficiently long, the intermediate hexatic phase completely vanished. With our data for the limited particle numbers we can not give a confident answer whether we observe a discontinuous transition in the present system.

\begin{figure}
\includegraphics{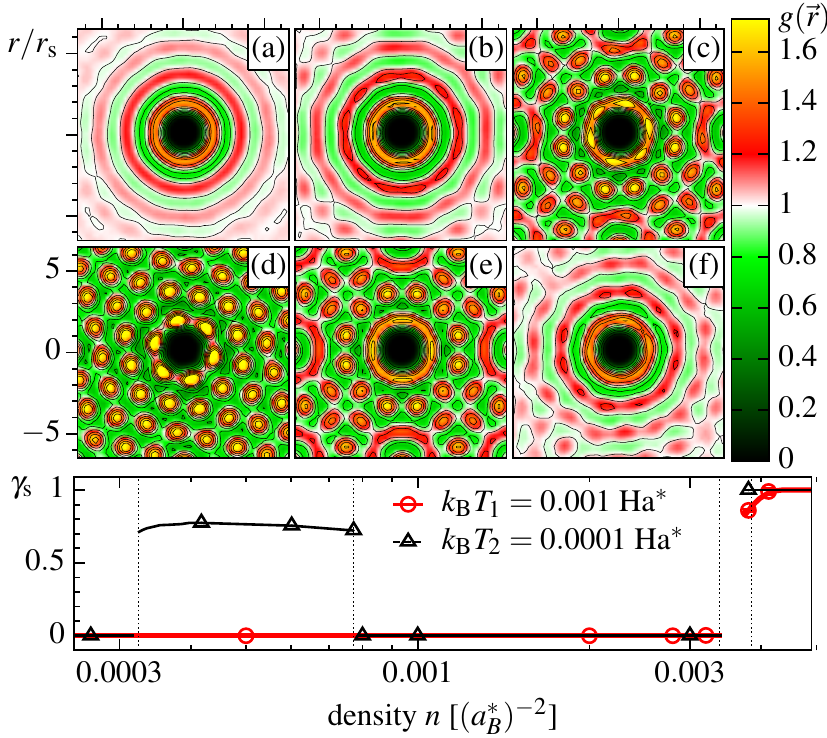}
\caption{(Color online) Isothermal freezing and melting of indirect excitons. a)-f): 2D PDF $g(\vec r)$ for $\kB T_1=0.001\,\Hart^*$ at densities $n\aB^{*2}$ of $0.84\cdot10^{-3}$ (a), $1.3\cdot10^{-3}$ (b), $1.7\cdot10^{-3}$ (c), $3.2\cdot10^{-3}$ (d),  $3.6\cdot10^{-3}$ (e), and $4.0\cdot10^{-3}$ (f). Bottom panel: Superfluid fraction $\gamma_\text{s}$, Eq.~\eqref{eq:sfwind}, vs. density for two temperatures. Symbols are PIMC results, lines are a guide to the eye. The increase of $\gamma_s$ at high density extends over a small finite range of solid-liquid coexistence which is due to the finite particle number in the simulations.}
\label{fig:meltdens}
\end{figure}

\subsection{Exciton quantum coherence. Superfluidity}

After analyzing emergence of \emph{spatial ordering} let us turn to the \emph{quantum coherence} properties of nonideal indirect excitons. In a 2D Bose system cooling leads to sudden emergence of coherence in the liquid phase -- the normal fluid -- superfluid transition. The phase boundary is governed by the Berezinskii-Kosterlitz-Thouless (KT) scenario~\cite{nelson_universal_1977}
and is given by the condition $\chi=4/\gs$ for the exciton quantum degeneracy parameter $\chi\equiv n\Lambda^2$
\begin{align}
  \kB T_\text{KT}(n_\text{s})=\frac{\pi}{2}\,n_\text{s}\,\frac{m_\elon^\inplane}{m_\exc}\Hart^*\;,
  \label{eq:bktest}
\end{align}
where $n_\text{s}=\gs n$ is the exciton superfluid density. Therefore, a key quantity is the superfluid fraction \gs, where $0\le\gs\le 1$. In PIMC simulations, it is directly computed from the statistics of the winding number $W$~\cite{ceperley_path_1995}:
\begin{align}
  \gs &=
\frac{m_\exc}{N\hbar^2\beta}\,\avr{W^2}\;, &
  \vec W = \sum_{i=1}^{N}\int_0^\beta\td t\, \frac{\td\vec r_i(t)}{\td t}
  \label{eq:sfwind}\;.
\end{align}
Typical simulation results for $\gs$ are shown in the bottom part of Fig.~\ref{fig:meltdens}.
\begin{figure}[t]
\vspace{-2.2cm}
\begin{center}
\centering\includegraphics[width=\columnwidth]{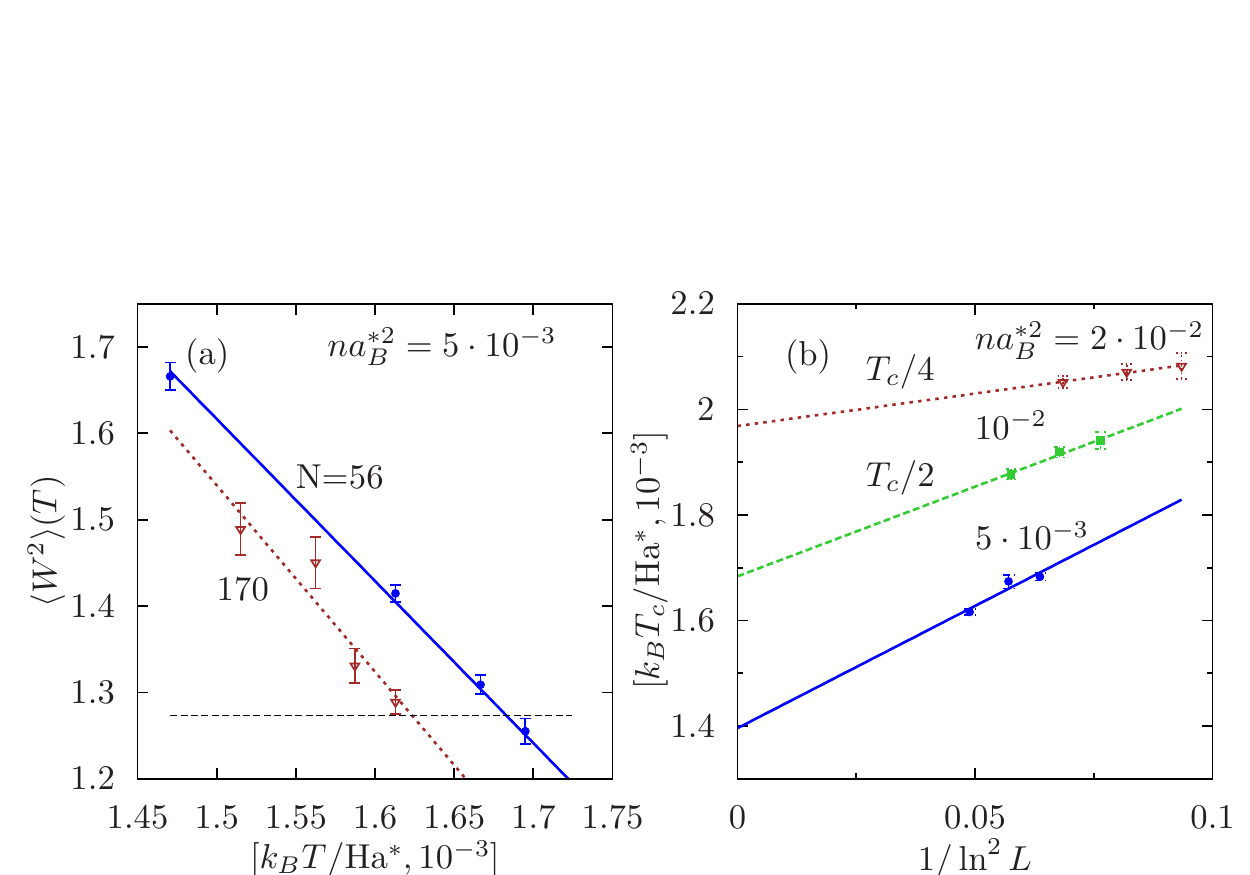}
\end{center}
\vspace{-0.5cm}
\caption{(a) Temperature dependence of the winding number $\langle W^2 \rangle(T)$ for the exciton numbers $N=56$ and $170$. Density $n \aB^{*2}=5 \cdot 10^{-3}$. The Berezinskii-Kosterlitz-Thouless transition temperature $T_\text{KT}(N)$ is determined by the condition,~\cite{filinov_berezinskii-kosterlitz-thouless_2010,nelson_universal_1977} $\langle W^2 \rangle(T_\text{KT})=4/\pi$, shown by the horizontal dashed line. (b) System size dependence of $T_\text{KT}(L)$ for three densities: $n \aB^{*2}=5 \cdot 10^{-3}, 10^{-2}$ and $2\cdot 10^{-2}$. Values of $T_\text{KT}$ are rescaled to fit into a single plot.}
\label{figs1}
\end{figure}

Figure \ref{figs1} illustrates the computation  of the winding number versus temperature (left) and the finite size scaling for the critical temperature $T_\text{KT}$ of the BKT transition (right). One observes a systematic shift of $T_\text{KT}(N)$ to lower values with an increase of the system size $N$. The extrapolation to the thermodynamic limit, $T_\text{KT}(L\rightarrow \infty)$, with $L=\sqrt{N/n}$, is obtained by fitting the simulation data by the equation $T_\text{KT}(L)=T_\text{KT}(\infty)+ b/\ln^2(L)$ . It is a direct consequence of the Kosterlitz-Thouless renormalization group analysis \cite{nelson_universal_1977} which is considered to be exact in the asymptotic regime of large $L$. This scaling allows us to make predictions for the phase transition line in a macroscopic system.

\subsection{Phase diagram of indirect excitons}

\begin{figure}
\includegraphics{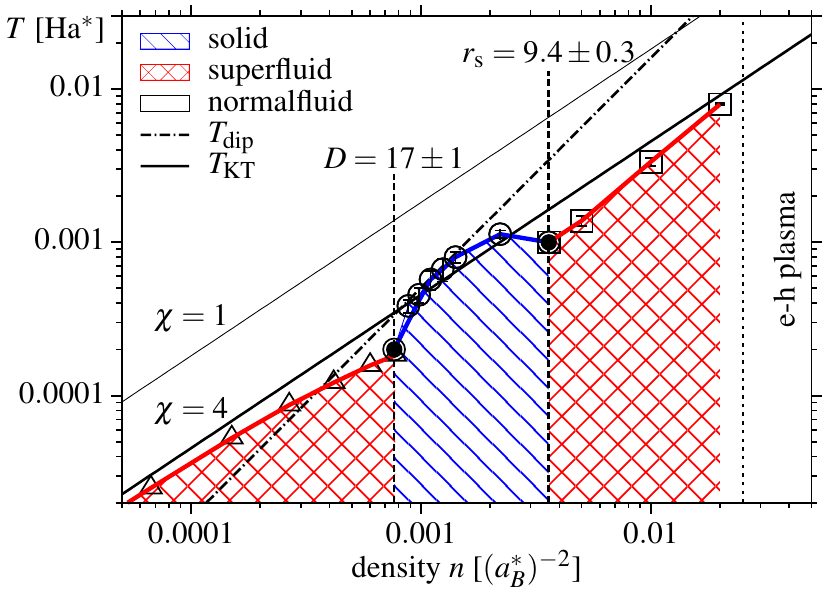}
\caption{(Color online) Phase diagram of 2D indirect excitons with $d=13.3\,\aB^*$. Circles and squares mark our PIMC results, data for triangles are from~\cite{filinov_berezinskii-kosterlitz-thouless_2010}. Vertical dashed lines ($D=17\pm1$ and $\rs=9.4\pm0.3$) indicate the two density induced quantum freezing (melting) transitions. Filled symbols mark the two triple points. The normal fluid--superfluid phase boundary is marked by the red line and is below the ideal estimate $T_\text{KT}$ according to Eq.~(\ref{eq:bktest}), cf. thick solid line labeled $\qdeg=4$. The line $T_\text{dip}$ marks the freezing transition of a classical 2D dipole system. The e-h plasma phase is beyond the present analysis.}
\label{fig:phase}
\end{figure}

We now summarize our findings in the complete phase diagram of indirect excitons in the density--temperature plane which is presented in Fig.~\ref{fig:phase}. The degeneracy line $\chi=1$ separates the regions of classical (above the line) and quantum  behavior (below). While classical excitons exist only in a fluid (or gas) phase the quantum region is composed of three different phases: a normal fluid, a superfluid and a crystal phase~\cite{note2}.
Correspondingly, there exist two triple points, at the upper left (right) edge of the crystal phase. At high temperature the excitons are in the fluid phase. Cooling leads either into the superfluid or crystal phase. There is no cooling transition from the superfluid to the crystal.

At low densities cooling always leads into the superfluid phase; the transition is accompanied by a sudden increase of \gs from zero to a finite value. The phase boundary is substantially below the upper limit $T_\text{KT}(n_\text{s}=n)$, Eq.~(\ref{eq:bktest}), and is in full agreement with our previous analysis for 2D dipoles~\cite{filinov_berezinskii-kosterlitz-thouless_2010} indicating that the exciton interaction is close to a dipole potential.
The picture suddenly changes when the density exceeds $n\aB^{*2} \approx 0.00078$:
the superfluid transition vanishes and, instead, a strong modulation of the PDF is observed signaling crystallization, cf. top row of Fig.~\ref{fig:melttemp}.
The critical density corresponds to a dipole coupling parameter~$D^\text{c}=17\pm1$ which agrees with studies of pure 2D dipole systems~\cite{buechler_strongly_2007, astrakharchik_quantum_2007}.
Note that the freezing temperature changes non-monotonically exhibiting a maximum value $T^\text{max}$ around $n\aB^{*2} \approx 0.002$.

The superfluid-solid transition is verified by simulating compression along several isotherms. At low temperature and low density, the superfluid fraction \gs starts from a high value until it suddenly drops to zero at the critical density $n\aB^{*2}\approx0.00078$, cf. bottom part of Fig.~\ref{fig:meltdens}. This behavior persists up to zero temperature, cf. Fig.~\ref{fig:phase}. Vanishing of quantum coherence upon crystallization is a general feature in this system and indicates that there is \emph{no supersolid phase of indirect excitons}.
If the temperature is above the left triple point the superfluid fraction is exactly zero, and compression leads to a phase transition from the normal fluid to the crystal phase, cf. $\gs$ for $T=0.001\,\Hart^*$ and the change of the PDF in Fig.~\ref{fig:meltdens}a--c.

\subsection{Reentrant quantum melting}
\label{reent_melt}
Interestingly, if the density is increased further, the exciton crystal melts, cf. Fig.~\ref{fig:meltdens}e,f, this time accompanied by a jump of the superfluid fraction from zero to about $0.9$. This indicates isothermal quantum melting to a (partially) superfluid exciton liquid. This occurs at a density of $n_\Coul\aB^{*2}=0.0036\pm0.0003$ corresponding to $\rs^\text{c}=9.4\pm0.3$ and, again, persist to zero temperature. At temperatures above the right triple point, $\kB T\gtrsim0.001\,\Hart^*$, melting and onset of superfluidity are decoupled: first the crystal melts into a normal fluid which becomes a superfluid only at a higher density, cf. Fig.~\ref{fig:phase}.

Thus the most striking feature of the exciton phase diagram is the existence of \emph{two quantum freezing (melting) transitions}, even in the ground state. At low-density excitons undergo \emph{pressure crystallization} which is characteristic for the behavior of dipole systems or, more generally, for neutral matter composed of atoms or molecules. In addition, at higher densities, there is a second transition: \emph{quantum melting} by compression. While such an effect is absent in conventional neutral matter it is ubiquitous in Coulomb systems, including the Wigner crystal of the strongly correlated electron gas, ion crystals in the core of white dwarf stars and nuclear matter in the crust of neutron stars.
The existence of this quantum melting transition in indirect excitons is due to the peculiar shape of the effective potential $V_{\exc\exc}$: one readily confirms in Fig.~\ref{fig:effpot} that at the critical density where the mean exciton-exciton distance equals $9.4\,\aB^*$,  $V_{\exc\exc}$ essentially follows the Coulomb repulsion of the holes (red dashed curve).

\begin{figure}
\includegraphics{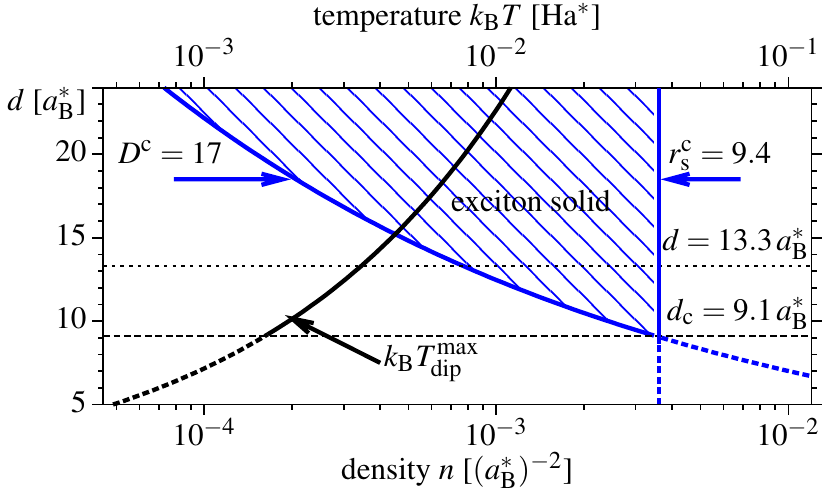}
\caption{Boundaries of the exciton crystal for different dipole moments~$d$. Lower abscissa: density range given by $D^\text{c}=17$ and $\rs^\text{c}=9.4$. Upper abscissa: maximum temperature estimated from $T^\text{max}_\text{dip}$. No solid phase exists for $d\le d_\text{c}\approx9.1\,\aB^*$.}
\label{fig:solid}
\end{figure}

\section{Conclusions}\label{final}

We have shown that a bosonic many-particle system possesses, besides its weakly nonideal Bose condensed gas and its superfluid liquid phases also a strongly correlated solid phase. Indirect excitons in semiconductor quantum wells have been found a favorable candidate due to their long-range pair interaction and the possibility to achieve strong nonideality by controlling the dipole moment with an external electric field. Based on first principle PIMC simulations we have computed the complete phase diagram in the region of the exciton crystal. (Quasi-)Long range crystalline order and macroscopic quantum coherence are found to be incompatible in an exciton crystal -- there is no supersolid phase, as long as the crystal is free of defects.

\subsection{Experimental realization}

The results presented above were computed for $d=13.3\,\aB^{*}$. Using values from Ref,~\cite{filinov_effective_2009} this dipole moment can be achieved in a ZnSe quantum well of width $L\approx50\,\text{nm}$ or a GaAs quantum well with $L\approx148\,\text{nm}$, both at an electric field strength of $E=20\,\text{kV/cm}$. The density interval for the exciton crystal is estimated as $1.3\cdot10^{9}\,\text{cm}^{-2}\ldots3.6\cdot10^{9}\,\text{cm}^{-2}$ for GaAs and $8.2\cdot10^9\,\text{cm}^{-2}\ldots3.8\cdot10^{10}\,\text{cm}^{-2}$ for ZnSe.
An estimate for the maximum temperature where the crystal can exist is obtained from the classical dipole melting curve,
\begin{align}
\kB T_\text{dip} &= c\,\frac{d^2}{\aB^{*2}}\,\left(n\aB^{*2}\right)^{3/2}\,\Hart^*\;,
\label{eq:tmax_dip}
\end{align}
where $c\approx0.09$,~\cite{kalia_interfacial_1981} and the critical density $n_\Coul\aB^{*2}=0.0036$ is being used. Taking into account that this value is approximately a factor $2$ too high, cf. Fig.~\ref{fig:phase}, we obtain the estimates $\kB T^\text{max}=0.17\,\text{K}$ (GaAs) and $\kB T^\text{max}=0.78\,\text{K}$ (ZnSe). These parameters are well within reach of current experiments. A particular advantage is that the upper density limit for exciton crystallization is a factor $16$ higher than the threshold for an electron Wigner crystal ($\rs\approx 37$). A suitable diagnostics for the excitonic crystalline phase can be Bragg scattering.~\cite{sperlich_electric_2009}

\subsection{Dependence of the quantum well width}\label{var_d}

Let us now analyze the dependence of the phase diagram on the dipole moment $d$. In semiconductor quantum wells the dipole moment can be varied in a broad range by varying the QW width or/and the electric field strength. As shown in Fig.~\ref{fig:solid}, an increase of $d$ reduces the lower density limit of the crystal phase whereas the upper boundary remains unchanged. Thus, the crystal phase expands with $d$, the maximum temperature $T^\text{max}$ grows quadratically, cf. Eq.~(\ref{eq:tmax_dip}) and Fig.~\ref{fig:solid}. Finally, there exists a minimum value $d_\text{c}=9.1\,\aB^*$ where the two limiting densities converge, and the exciton crystal phase vanishes.

\subsection{Outlook}

Let us now briefly discuss effects which have been neglected by the present model, most importantly, disorder and thermal relaxation.

To reduce the effect of the exciton localization at surface imperfections we considered the model of a single wide QW ($L > 400$ \AA). This allows us to completely neglect the effect of 1 monolayer well width fluctuations on the exciton binding energy and localization. Some quantitative analysis can be found in A. Filinov \emph{et al.}~\cite{exc_loc} In our case, the in-plane size of the exciton wavefunction is comparable to the dipole moment $d=13.3 \aB^* \approx 400$ \AA and is, therefore, of the order of the characteristic lateral size of the interface fluctuations $\sim 400$ \AA \, (see D. Gammon \emph{et al.}~\cite{gammon}). Hence, once the exciton is on the top of the defect, the corresponding potential gets significantly smoothed.

In many optical experiments excitons are created in a highly non-equilibrium state with a possible coherence and coupling to the laser field. Such conditions, certainly, complicate both the interpretation of the experiment and the theoretical description, and have been studied in detail for polaritons. In contrast, we consider an experimental realization, where the excitons are created by an optical pulse, which is switched off after a short duration, or is  periodically repeated with a delay of several microseconds, sufficient for the exciton equilibration. Fast exciton recombination is prevented by the spatial e-h separation due a constantly applied electric field. This situation is experimentally feasible as was shown by Z.~Vörös \emph{et al.}~\cite{vobs}.

Finally, the most striking feature of the crystal of indirect excitons, confirmed by the simulations, is two quantum melting transitions which persist at zero temperature: at low densities it melts by expansion whereas at high densities it melts when being compressed. The origin of this unusual and rich phase diagram has been traced to the non-trivial form of the exciton interaction potential.
With it the exciton solid combines features of conventional neutral matter (exhibiting crystallization by compression) and Coulomb matter (quantum melting by compression), as found for instance in exotic compact stars.

\section{Acknowledgements}

We thank D. Hochstuhl for performing multiconfiguration Hartree-Fock calculations for the exciton interaction energy. Stimulating discussions with Yu. Lozovik and P. Ludwig and financial support by the Deutsche Forschungsgemeinschaft (project FI 1252/1 and SFB-TR24 project A5) are gratefully acknowledged.

\nocite{bonitz_crystallization_2005, laikhtman03}

\bibliography{exc11}

\begin{thebibliography}{43}
\bibitem[{not({\natexlab{a}})}]{note1}
\bibinfo{note}{An alternative realization are two coupled $n-$ and $p-$doped
  semiconductor layers.}
\bibitem[{not({\natexlab{d}})}]{note4}
\bibinfo{note}{Spin degree of freedom are omitted in the present analysis, as
  this requires a significantly more elaborated simulations. The model used for
  the exciton interaction potential, is assumed to have a significantly larger
  effect on the results, when the spin fluctuations in the ferromagnetic
  phase~\cite{laikhtman03}.}
\bibitem[{not({\natexlab{f}})}]{note6}
\bibinfo{note}{As we consider low temperatures ($T \lesssim 0.01$~Ha$^*$) for
  the electrons, only the ground state solution is included. This is due to the
  large energy gap to the first excited state}.
\end{thebibliography}

\end{document}